\documentclass{article}
\usepackage[english]{babel}
\usepackage[T2A]{fontenc}

\usepackage[tbtags]{amsmath}
\usepackage{amsfonts,amssymb,mathrsfs,amscd}
\usepackage{tikz}
\usepackage{hyperref}
\usepackage{lscape}

\numberwithin{equation}{section}

\begin{document}
    \title{Mathematical aspects of space-time\\ horizontal ray method}
    \author{A. Kaplun\footnote{University of Haifa
,\,\,\href{mailto: alex.v.kaplun@gmail.com}{alex.v.kaplun@gmail.com},\,\,\href{mailto: a.kaplun.work@gmail.com}{a.kaplun.work@gmail.com}, supported by the Center for Integration in Science, Israel Ministry of Aliyah and Integration and post-doctoral scholarship at the Leon H. Charney School of Marine Sciences}, B. Katsnelson\footnote{University of Haifa
,\,\,\href{mailto: bkatsnels@univ.haifa.ac.il}{bkatsnels@univ.haifa.ac.il}, supported by ISF-946/20}}
\date{}

    \maketitle

\begin{abstract}
The following development of the well-known ”vertical modes and horizontal rays” approach for acoustic waves propagation in shallow water, introduced in works \cite{burridge2005horizontal,weinberg1974horizontal,connor1974complex}, is studied. In this approach we study so-called space-time horizontal rays \cite{babich1998space}, constructed on the base of  decomposition of the sound field, depending on time, over adiabatic vertical modes (solutions of the Sturm-Liouville problem). Using this technique we obtain different properties of signals, propagating in underwater waveguide, such as space-time caustics \cite{kravtsov2012caustics}, and provide rather simple method for the prediction of the form of the signal and all its parameters (amplitude and frequency modulation, different front angles, etc.) at some point of observation.
\end{abstract}
\section{Introduction}
This paper aims to fill the gap existing between two approaches to the solution of the problem of acoustic wave propagation in a shallow sea. On the one hand, there is a representation for the solution of the wave equation 
\begin{align}
    & \Delta u - \frac{1}{c^2}\frac{\partial^2  }{\partial t^2} u  = 0
\end{align}
in the form of decomposition by vertical modes - eigenfunctions along the vertical axis\cite{burridge2005horizontal, weinberg1974horizontal, felsen1981hybrid}:
\begin{align}
    & u(t,\mathbf{r},z)\sim \int \sum_{n} a_n(\mathbf{r},\omega) \psi_n (z,\mathbf{r},\omega)\exp(- i \omega t) d \omega.
\end{align}
This approach provides an adequate description of the field under conditions of frequency variation in the small range $[\omega_0 - \delta\omega,\omega_0 + \delta\omega]$, but obtaining results in the case of substantial frequency modulation requires multiple solutions of the equations for different modes and frequencies, followed by application of the Fourier transform \cite{katsnelson2018variability,katsnel2012space}. Therefore, it is difficult and resource-consuming to apply this method for numerical calculation of wave process parameters in the general case. On the other hand, there is a method of consideration of wave processes in the presence of frequency dispersion and modulation - space-time rays \cite{connor1974complex,kirpichnikova1983reflection,bergman2005generalized,vcerveny1982space,babich1980space,babich1984complex} used in various fields of mathematics and physics. Such an object has previously been defined \cite{weinberg1974horizontal} for the study of horizontal refraction, but further development of the mathematical apparatus is required to study and predict the various observed effects (other than refraction proper).The particular feature of the proposed approach is also the fact that all observed effects are determined by the structure of the function $q_l$ (eigenvalue corresponding to the vertical mode), which contains all necessary information about possible dispersion effects, therefore it does not require the construction of a separate model for each specific case (parameters of sound speed, bottom, obstacles, etc.). Detailed reviews of the two methods described above are given in \cite{babich1998space,katsnelson2012fundamentals,etter2018underwater}. It will also be important to note that the approach we consider is close to the one used to solve the wave equation in a material medium with arbitrary dispersion \cite{kravtsov1990geometrical}:
\begin{align}
& \Delta u - \frac{1}{c^2}\frac{\partial^2 }{\partial t^2} \check{M}[u] = 0,
\end{align}
where $\check{M}$ is some integral operator describing the properties of the medium.

An additional motivation for this work is the fact that the developed mathematical apparatus turns out to be convenient for describing wave processes in other dispersing media as well. Several effects due to the influence of the dispersion in an inhomogeneous medium have been found in the study of electromagnetic wave propagation: in the optics of short, picosecond and femtosecond pulses \cite{topp1975group,bor1985group,hebling1996derivation,osvay2004angular,hebling2008generation}. In particular, these effects manifest themselves in the form of a tilt between the phase front and the pulse front (pulse front tilt) after passing a dispersing medium (prism) \cite{topp1975group,bor1985group,hebling1996derivation,osvay2004angular} and are used in nonlinear generation of terahertz pulses \cite{hebling2008generation}.
\section{Vertical modes and field distribution}
\subsection{Wave equation}
Let us consider the wave equation describing the propagation of acoustic waves in the sea
\begin{align}
    & \left[\frac{\partial ^2}{\partial X^2} + \frac{\partial ^2}{\partial Y^2} + \frac{\partial ^2}{\partial Z^2} - \frac{1}{c(X,Y,Z)^2} \frac{\partial ^2}{\partial T^2}\right] u(T,X,Y,Z) = 0.
\end{align}
Here $X,Y$ are horizontal coordinates and $Z\ge 0$ is vertical, with $Z=0$ being the equation of the water surface, $Z= h(X,Y)$ being the equation of the lower surface (bottom) at the point $(X,Y)$, and $c(X,Y,Z)$ being the speed of sound in the medium, which depends on the depth $Z$ significantly more than on the horizontal coordinates $(X,Y)$ \cite{weinberg1974horizontal}. The desired function $u$ has the physical meaning of acoustic pressure. Let us rewrite the equation in the form
\begin{align}
    & \left[- n^2(X,Y,Z) \frac{\partial ^2}{\partial (c_0 T)^2} +\frac{\partial ^2}{\partial X^2} + \frac{\partial ^2}{\partial Y^2} + \frac{\partial ^2}{\partial Z^2} \right] u(c_0 T,X,Y,Z) = 0,
\end{align}
where $c_0$ is some typical speed of wave propagation, and the following condition is satisfied
\begin{align}
    & n(X,Y,Z) := \frac{c_0}{c(X,Y,Z)}\sim 1.
\end{align}
The boundary conditions for the wave equation are written in the form \begin{align}
    & u(T,X,Y,Z)|_{Z=0} = 0,\\
    & u (T,X,Y,Z)|_{Z=h(X,Y)+} = u (T,X,Y,Z)|_{Z=h(X,Y)-}\\
    & \frac{1}{\rho^{+}} \frac{\partial u (T,X,Y,Z)}{\partial Z}|_{Z=h(X,Y)+}  = \frac{1}{\rho^{-}} \frac{\partial u (T,X,Y,Z)}{\partial Z}|_{Z=h(X,Y)-}.
\end{align}
where $\rho^{+},\rho^{-}$ are the densities of the matter above and below the bottom, respectively, and we also require that the radiation conditions for the function $u$ are satisfied. We will assume that the following estimates relating the derivatives for different variables are satisfied:
\begin{align}
    & \frac{\partial}{\partial (\varepsilon X)}  \sim \frac{\partial}{\partial (\varepsilon Y)}  \sim \frac{\partial}{\partial (\varepsilon c_0 T)} \sim \frac{\partial}{\partial Z}.
\end{align}
Let us introduce new notations reflecting these properties:
\begin{align}
    & \vec{r} = (x,y) := (\varepsilon X,\varepsilon Y),\quad \tau:= \varepsilon c_0 T,\quad z := Z,\\
    & \frac{\partial}{\partial x}  \sim \frac{\partial}{\partial y}  \sim \frac{\partial}{\partial \tau } \sim \frac{\partial}{\partial z},
\end{align}
The wave equation is rewritten in the form
\begin{align}
    & \left[- n^2(x,y,z) \frac{\partial ^2}{\partial \tau^2} +\frac{\partial ^2}{\partial x^2} + \frac{\partial ^2}{\partial y^2} + \varepsilon^{-2}\frac{\partial ^2}{\partial z^2} \right] u(\tau,x,y,z) = 0.
\end{align}
In real physical applications, it can be assumed that $\varepsilon<0.1$. Let us take as an ansatz\footnote{Choosing such ansatz we consider a problem of the propagation of only one vertical mode and ommit effects related to mode coupling.} for the solution of such equation the following one:
\begin{align}
    & u (\tau,x,y,z) =  A(\tau,x,y)\psi(\tau,x,y,z)\exp{\left\{i\varepsilon^{-1}\phi(\tau,x,y)\right\}},
\end{align}
where
\begin{align}
    %&\kappa_l = \kappa_l(\tau,X,Y):= \frac{\partial \phi_l}{\partial \tau},\\
    & A(\tau,x,y) := \sum_{j=0}^{\infty} (i\varepsilon)^j A^j(\tau,x,y).
\end{align}
Substitution of the ansatz after the reduction of the exponential multiplier leads to equality:
\begin{align}
    \notag& A \varepsilon^{-2}\bigg[\frac{\partial^2 \psi}{\partial z^2} + \left\{n^2 \left(\frac{\partial \phi}{\partial\tau}\right)^2  - \left(\vec{\nabla\phi}\right)^2\right\}\psi\bigg] + \\ \notag&i\varepsilon^{-1}\psi\bigg[A\left\{\Delta\phi -n^2 \frac{\partial^2\phi}{\partial\tau^2}\right\}  + 2 \left(\vec{\nabla} \phi,\vec{\nabla}A\right)-2n^2\frac{\partial\phi}{\partial\tau}\frac{\partial A}{\partial\tau}\bigg] +\\ \notag&+ 2i\varepsilon^{-1}A\bigg[{ -n^2\frac{\partial \phi}{\partial\tau}\frac{\partial \psi}{\partial\tau}}+ \left(\vec{\nabla} \phi,\vec{\nabla}\psi\right)\bigg] +  \\ &
   + \bigg[\Delta A \psi + A\Delta\psi - n^2{ A\frac{\partial^2\psi}{\partial\tau^2}}+ 2 \left(\vec{\nabla} A,\vec{\nabla}\psi\right) \bigg] = 0
\end{align}
The gradient $\vec{\nabla}$ and the Laplace operator $\Delta$ are the operators with respect to the variables $x$ and $y$. Let us define the instantaneous frequency and the instantaneous wave vector using the equations:\begin{align}
    k_0:= \frac{\partial \phi}{\partial \tau},\quad \vec{k}:= \frac{\partial \phi}{\partial \vec{r}} = \left(\frac{\partial \phi}{\partial x},\frac{\partial \phi}{\partial y}\right)=(k_x,k_y).
\end{align}
Then the first two equations from decomposition by powers of $\varepsilon$ have the form:
\begin{align}
    \label{first order}& \bigg[\frac{\partial^2 \psi}{\partial z^2} +\left\{ n^2 k_0^2  - |\vec{k}|^2\right\}\psi\bigg]A^0 = 0,\\
    \notag& \bigg[A^0\left\{\Delta\phi -n^2\frac{\partial^2\phi}{\partial\tau^2}\right\}  +2 \left(\vec{\nabla} \phi,\vec{\nabla}A^0\right)-2n^2\frac{\partial\phi}{\partial\tau}\frac{\partial A^0}{\partial\tau}\bigg]\psi\,\,+\\ \label{second order}& + 2A^0\bigg[{ -n^2\frac{\partial \phi}{\partial\tau}\frac{\partial \psi}{\partial\tau}}+ \left(\vec{\nabla} \phi,\vec{\nabla}\psi\right)\bigg] = 0.
\end{align}
\subsection{Vertical modes}
Consider a family of Sturm-Liouville operators depending on $\vec{r},k_0$ as parameters:
\begin{align}
    & L[\vec{r},k_0]\psi:= \left[\frac{d^2}{d z^2} + n^2(\vec{r},z) k_0^2\right] \psi\\
    & \psi_{z=0} = 0,\quad \psi|_{z = h(\vec{r})+} = \psi|_{z = h(\vec{r})-},\\ 
    & \frac{1}{\rho(\vec{r})^{+}} \frac{\partial \psi}{\partial z}|_{z = h(\vec{r})+}  = \frac{1}{\rho(\vec{r})^{-}} \frac{\partial\psi}{\partial z}|_{z = h(\vec{r})-},
\end{align}
and also $\psi$ satisfies the radiation conditions at infinity\footnote{In our case it means $\psi\in L_2(\mathbb{R}_+)$.}. Let us assume that such operators have a finite positive discrete spectrum \begin{align}
&\lambda_0(\vec{r},k_0)>\lambda_1(\vec{r},k_0)>\dots>\lambda_{n(\vec{r},k_0)}(\vec{r},k_0)>0,
\end{align}  consisting of simple eigenvalues, and the strict inequalities are satisfied for the whole range of variation of the parameters $\vec{r},k_0$ (the discrete positive spectrum has multiplicity $1$). We will refer to the eigenfunctions $\psi_0,\dots,\psi_{n(\vec{r},k_0)}$ corresponding to each spectrum as vertical modes. Due to the positivity of $\lambda$, we introduce notation:
\begin{align}
& q_l (\vec{r},k_0) := \sqrt{\lambda_l (\vec{r},k_0)} >0.
\end{align}
We can define the scalar product between the eigenfunctions by the following formula
\begin{align}
    & \langle \psi_l,\psi_{l'}\rangle := \rho(\vec{r})^{+}\int_{0}^{h(\vec{r})}\psi_l\psi_{l'} dz + \rho(\vec{r})^{-}\int_{h(\vec{r})}^{\infty}\psi_l\psi_{l'} dz = \delta_{l l'}.
\end{align}
The general properties of such operators are a separate object of study (see \cite{katsnelson2012fundamentals}); here we will only use known results. 
\subsection{Eikonal equation}
Let us consider in detail the equation \ref{first order}. Let $\psi = \psi_l$ be some fixed vertical mode. From this equation we obtain (for arbitrary amplitude $A^0$) a nonlinear equation on the function $\phi$ (here and below we omit the index $l$ and assume that all considerations refer to the fixed mode): 
\begin{align}
    &  q^2 \left(\vec{r},\frac{\partial \phi}{\partial \tau}\right) - |\vec{\nabla} \phi|^2   = 0.
\end{align}
We will refer to this equation as the eikonal equation. It can be solved using the method of characteristics. Let us introduce the Hamiltonian $\mathcal{H}$ by the following formula
\begin{align}
    & \mathcal{H} (k_0,\vec{k},\vec{r}) := |\vec{k}|^2 - q^2(\vec{r},k_0)\equiv 0.
\end{align}
Then the space-time rays (characteristics) are defined as solutions of the canonical Hamiltonian system \cite{babich1998space}:
\begin{equation}\label{ham syst}
    \frac{d \hat{r}}{d\xi} = \frac{\partial\mathcal{H}}{\partial \hat{k}},\quad 
    \frac{d\hat{k}}{d\xi} = -\frac{\partial\mathcal{H}}{\partial \hat{r}}, 
\end{equation}
where $\xi$ is some one-dimensional parameter along the ray, which has no explicit physical meaning, and $\hat{\cdot}$ denotes objects referring to the three-dimensional space $(\tau,\vec{r})$. For such Hamiltonian the system has the form:
\begin{align*}
    & \frac{d \hat{r}}{d \xi} = 2 q \left(-\frac{\partial q}{\partial k_0} , \frac{k_x}{q},\frac{k_y}{q}\right) = 2 q \left(-\frac{\partial q}{\partial k_0} , \frac{\vec{k}}{|\vec{k}|}\right),\\
    & \frac{d \hat{k}}{d \xi} = 2 q \left(\frac{\partial q}{\partial \tau} , \frac{\partial q}{\partial x},\frac{\partial q}{\partial y}\right) = 2 q \left(0 , \vec{\nabla}q\right).
\end{align*}
Hence we see that the system has a natural parameter along the ray defined by the equality
\begin{align}
    & d s := 2 q d \xi.
\end{align}
Let us introduce the notations
\begin{align}
   \label{vec kappa def} & \hat{\kappa} :=  \left(-\frac{\partial q}{\partial k_0},\frac{k_x}{q},\frac{k_y}{q}\right) = \left(\kappa_0, \vec{\kappa}\right) , \quad|\vec{\kappa}| =1,\\
    & v:= (\kappa_0)^{-1}  ,\quad   |\hat{\kappa}|^2 = 1 + \kappa_0^2 = \frac{1 + v^2}{v^2}= 1 +  \left(\frac{\partial q}{\partial k_0}\right)^2.
\end{align}
Defining $\kappa_0$ and $v$ simultaneously is convenient because $v$ represents the group velocity of wave propagation. Then for the parameter $s$ we have 
\begin{align}
   \label{HJ eq orig} & \frac{d \hat{r}}{d s} = \hat{\kappa},\quad \frac{d \hat{k}}{d s} = \left(0 , \vec{\nabla}q\right).
\end{align}
Hence we can see the physical meaning of the parameter $s$ - $ds$ is the element of length along the spatial ($\vec{r}$) part of the ray. Note also that the vectors $q \hat{\kappa}$ and $\hat{k}$ differ only in the first coordinate: in the first one it is $- q \frac{\partial q}{\partial k_0}$, and in the second one it is $k_0$.
\subsection{Transport equation}
Now we will consider the equation \ref{second order}. Let us multiply it by $A^0$ and rewrite it in a more convenient form (omitting the index $0$):
\begin{align}
\notag& 
\bigg[A^2\left\{\Delta\phi -n^2\frac{\partial^2\phi}{\partial\tau^2}\right\}  +2 A\left(\vec{\nabla} \phi,\vec{\nabla}A\right)-2n^2A\frac{\partial\phi}{\partial\tau}\frac{\partial A}{\partial\tau}\bigg]\psi+\\ \label{trans eq}& + 2A^2\bigg[{ -n^2\frac{\partial \phi}{\partial\tau}\frac{\partial \psi}{\partial\tau}}+ \left(\vec{\nabla} \phi,\vec{\nabla}\psi\right)\bigg]= 0
\end{align}
It is not difficult to see that the dependence on the parameter $z$ is contained only in the functions $m$ and $\psi$. Consider the auxiliary equation \begin{align}
    & \langle\psi'',\psi\rangle + k_0^2 \langle n^2 \psi,\psi\rangle = q^2 \langle\psi,\psi\rangle,
\end{align}
obtained by scalar multiplication of the equation $L\psi = q^2 \psi$ by the same eigenfunction $\psi$. It shows that
\begin{align}
    & \langle n^2\psi,\psi\rangle = \frac{1}{k_0^2} (q^2 + \langle\psi',\psi'\rangle)\approx \frac{q}{k_0} \frac{\partial q}{\partial k_0}.
\end{align}
Let us also compute expressions for the other scalar products that we will need:
\begin{align}
    & \langle (\vec{\nabla\phi},\vec{\nabla\psi}),\psi\rangle = \frac{1}{2} (\vec{\nabla} \phi,\vec{\nabla}\langle\psi,\psi\rangle) = 0,\\
        & \langle n^2\frac{\partial \psi}{\partial \tau},\psi\rangle = \frac{1}{2}\frac{\partial}{\partial\tau} \langle n^2\psi,\psi\rangle\approx \frac{1}{2}\frac{\partial}{\partial\tau}\left[\frac{q}{k_0}\frac{\partial q}{\partial k_0}\right].
\end{align}
Then, taking the scalar product of the transfer equation \ref{trans eq} with the function $\psi$, we have
\begin{align}
&  \frac{\partial}{\partial \tau} \left[-A^2 k_0\langle n^2\psi,\psi\rangle\right]+\vec{\nabla}\cdot \left[A^2\vec{k}\right]\langle\psi,\psi\rangle = 0
\end{align}
Using the approximation for $\langle n^2\psi,\psi\rangle $, we obtain the final form for the transport equation of the principal component of the amplitude
\begin{align}
    &   \frac{\partial}{\partial \tau} \left[-A^2 q \frac{\partial q}{\partial k_0}\right]+\vec{\nabla}\cdot \left[A^2\vec{k}\right]= 0.
\end{align}
Comparing to the definition of the vector $\hat{\kappa}$ \ref{vec kappa def}, we notice that this equation can also be written in the divergence form
\begin{align}
    \label{trans eq div} & \hat{\mathrm{div}} \left(qA^2\hat{\kappa}\right) = 0,
\end{align}
where $\hat{\cdot}$ again denotes operations in three-dimensional space $(\tau,\vec{r})$. A more precise equation can also be written in the divirgence form
\begin{align}
    & \hat{\mathrm{div}} \left(qA^2\hat{\kappa}'\right) = 0,&& \hat{\kappa}'=\left(-\frac{k_0}{q}\langle n^2\psi,\psi\rangle,\frac{\vec{k}}{q}\right)^T,
\end{align}
however, in the approximation of small dependence of the eigenfunctions on the frequency, $\vec{\kappa}'\approx\vec{\kappa}$ is fulfilled.
\subsection{Hamilton system}
Let us consider in more detail the equations of the rays \ref{HJ eq orig}. Let us use the equality $\vec{k} = q \vec{\kappa}$ and derive\begin{align}
    & \frac{d\vec{k}}{d s} = q\frac{d\vec{\kappa}}{d s} + \vec{\kappa}\, \frac{dq}{d s}= q\frac{d\vec{\kappa}}{d s} +   (\vec{\nabla}q,\vec{\kappa})\vec{\kappa}.
\end{align}
Hence
\begin{equation}\label{eq for vec l}
    \frac{{d}\vec{\kappa}}{{d} s} = \frac{\vec{\nabla}q}{q} -  \left(\frac{\vec{\nabla}q}{q},\vec{\kappa}\right)\vec{\kappa}.
\end{equation}
Let $\alpha$ be the angle between the vector $\vec{\kappa}$ and the axis $x$. Then
\begin{equation}
    \vec{\kappa} (s) =\left(
    \cos \alpha (s),
    \sin \alpha (s)
   \right)^T. 
\end{equation}
Let us define $2\times 2$ - matrix $J$:
\begin{equation}
    J:=\left(\begin{array}{cc}
         0 & -1 \\
         1 & 0
    \end{array}\right),\quad J^2 =\left(\begin{array}{cc}
         -1 & 0 \\
         0 & -1
    \end{array}\right)= - I.
\end{equation}
Then the equation \ref{eq for vec l} can be replaced by the equation for angle $\alpha$:
\begin{equation}
    \frac{{d}\alpha}{{d} s} =\left( \frac{\vec{\nabla}q}{q} ,J\vec{\kappa}(\alpha)\right).
\end{equation}
Let us introduce the angle $\beta$, which allows us to write the vector $\hat{\kappa}$ in a more illustrative form:
\begin{align}
    \label{def beta}&     & \tan \beta := \kappa_0 = \frac{1}{v},\quad |\hat{\kappa}| = \frac{1}{\cos \beta} ={\sqrt{1 + \kappa_0 ^2 }} = \frac{\sqrt{1+v^2}}{v}  .
\end{align}
Then
\begin{align}
    & \hat{\kappa} = \frac{1}{\cos\beta} \left(\begin{array}{c}
         \sin \beta  \\
         \cos\alpha \cos \beta\\
         \sin \alpha \cos \beta
    \end{array}\right) = \left(\begin{array}{c}
         \tan \beta  \\
         \cos\alpha \\
         \sin \alpha 
    \end{array}\right).
\end{align}
It is easy to see that the angles $\alpha$ and $\beta$ have a clear physical meaning: $\alpha$ describes the deviation of the spatial ray from a certain reference direction (in our case, from the $x$-axis), and $\beta$ is related to the group velocity and motion of the ray in the temporal part of spacetime. The definition of the angle $\beta$ for an arbitrary medium provides a fair analog of Snellius's law: 
\begin{align}
    & v \tan \beta = 1,
\end{align}
where $v$ is the group velocity. 

Thus, space-time rays are determined by a system of equations:
\begin{align}
    & \frac{{d}\tau}{{d}s} = \kappa_0 ,  &&\frac{{d}\kappa_0}{{d}s} = 0,\\
    &\frac{{d}\vec{r}}{{d}s}= \vec{\kappa}(\alpha)  ,&&     \frac{{d}\alpha}{{d}s} =  \left(\frac{\vec{\nabla}q}{q},J\vec{\kappa}(\alpha)\right),
\end{align}
or, in case if it is more convenient to consider $\tau$ as a parameter along the ray, 
\begin{align}
    & \frac{{d}s}{{d}\tau} = v,  &&\frac{{d}\kappa_0}{{d}\tau} = 0,\\
    &\frac{{d}\vec{r}}{{d}\tau} = v\vec{\kappa}(\alpha)  ,&&     \frac{{d}\alpha}{{d}\tau} =  v\left(\frac{\vec{\nabla}q}{q},J\vec{\kappa}(\alpha)\right).
\end{align}

\subsection{Transport equation and phase change along ray}
Let us now take a closer look at the transport equation \ref{trans eq div} in divergence form:
\begin{align}
& \hat{\mathrm{div}} \left(qA^2\hat{\kappa}\right) = 0,
\end{align}
Considering an infinitesimal ray tube along the space-time ray (in the direction of the three-dimensional vector $\hat{\kappa}$), we obtain the invariance of the expression
\begin{align}
    & q A^2 |\hat{\kappa}| d \mathcal{S} = A^2 q\sqrt{1+\kappa_0^2} { }d \mathcal{S} \equiv \mathrm{const},
\end{align}
where $d \mathcal{S}$ is the cross-sectional element of this tube. Hence we obtain that, for the amplitude along the ray the following formula is valid
\begin{equation}\label{a(s) first}
    A(\tau) = A(0) \times \sqrt{\frac{g(0)}{g(\tau)}} \times \sqrt{\frac{d\mathcal{S}(0)}{d\mathcal{S}(\tau)}}, \quad  g(\tau):=\frac{q(\tau)}{\sqrt{1+\left(\frac{\partial q}{\partial k_0}\right)^2}}.
\end{equation}
An explicit expression can be obtained for the ratio of the cross section elements of a ray tube using ray coordinates. Let $\mu,\nu$ be such that $\tau,\mu,\nu$ form a set of ray coordinates for which the following is true 
\begin{align}
    & D(s,\mu,\nu):=\left|\det\mathcal{J}\right| \neq 0
\end{align}
in some neighborhood of the initial point $s_0,\mu_0,\nu_0$, where \begin{align}
    & \mathcal{J}:= \frac{\partial(\tau,x,y)}{\partial (s,\mu,\nu)}
\end{align}
- is the Jacobi matrix, and $D$ is the Jacobian of the transition to ray coordinates. Then the formula \ref{a(s) first} can be written in the following form:
\begin{equation}\label{a(s) second}
    A(\tau) = A(0) \times \sqrt{\frac{g(0)}{g(\tau)}} \times \sqrt{\frac{D(0)}{D(\tau)}}.
\end{equation}
Now let us consider the change of the phase $\phi$ along the ray. From the definition of the wave vector $\hat{k}$, it follows that
\begin{equation}
    d \phi  = k_0d\tau + (\vec{k},d\vec{r}) = (\hat{k},\hat{\kappa})ds = v(\hat{k},\hat{\kappa})d\tau.
\end{equation}
Using explicit form of the vectors $\hat{k},\hat{\kappa}$, we obtain
\begin{equation}
     d \phi = \left(q - k_0 \frac{\partial q}{\partial k_0}\right) ds   = k_0 d\tau  -\frac{v}{v_{\phi}}k_0d\tau = (k_0 - qv)d\tau,
\end{equation}
where $v_{\phi}:=-\frac{k_0}{q}$ and $v$ are the phase and group velocities, respectively. Then 
\begin{equation}
    \phi(\tau) = \phi(0) + k_0\tau + \int_{0}^{\tau}q(\tau')v(\tau')d\tau'.
\end{equation}
Note that the phase remains constant for any ray on which the phase and group velocities are identical. 

\section{Signal propagation}
\subsection{General equations}
Let us consider the following formulation of the problem. Let $\rho$ denote time as an observable quantity, and $\tau$ denote time as a parameter along the ray. As ray coordinates, we consider the triple $(\tau,\nu,\mu)$ defining the rays using the following relations:
\begin{align}
    & \frac{{d}\rho}{{d}\tau} = 1,  &&\frac{{d}\vec{r}}{{d}\tau} = v\vec{\kappa}(\alpha) ,\\
    &\frac{{d}k_0}{{d}\tau} = 0 ,&&     \frac{{d}\alpha}{{d}\tau} =  v\left(\frac{\vec{\nabla}q}{q},J\vec{\kappa}(\alpha)\right),\\
    & \rho(\mu, \nu)|_{\tau =0} = {\rho_0}(\mu,\nu), && \vec{r}( \alpha_0, \xi)|_{\tau =0} =\vec{r}_0(\mu,\nu), \\
    & k_0 (\mu, \nu)|_{\tau =0} = k_0 (\mu,\nu), && \alpha ( \mu, \nu)|_{\tau =0} = \alpha_0(\mu,\nu).
\end{align}
Now let's consider the variation of beam parameters under small changes of initial data
\begin{align}
  \mu := \mu_0 + \delta \mu,\quad \nu' := \nu + \delta\nu.
\end{align}
Consider two close vectors:
\begin{align}
    \vec{r} := \vec{r}(\tau,\mu,\nu),\quad \vec{r}' := \vec{r}(\tau,\mu',\nu').
\end{align}
The following representation is valid: 
\begin{align}
    \vec{r}' =\vec{r} + \Delta_{||}(\tau,\mu,\nu) \vec{\kappa}(\alpha(\tau,\mu,\nu)) + \Delta_{\perp}(\tau,\mu,\nu) J\vec{\kappa}(\alpha(\tau,\mu,\nu)),
\end{align}
where it is assumed that $\Delta_{|||}$ and $\Delta_{\perp}$ depend in some way on $\delta \mu$ and $\delta \nu$. We define $\delta \vec{r}$ by the equality:
\begin{align}
    \delta \vec{r} := \vec{r}' - \vec{r}= \Delta_{||}(\tau,\mu,\nu) \vec{\kappa}(\alpha(\tau,\mu,\nu)) + \Delta_{\perp}(\tau,\mu,\nu) J\vec{\kappa}(\alpha(\tau,\mu,\nu)).
\end{align}
Differentiating with respect to $\tau$, we obtain
\begin{align}
    & \frac{{d}\delta \vec{r}}{{d}\tau} = \frac{{d}\Delta_{||}\vec{\kappa} }{{d}\tau} + \frac{{d}\Delta_{\perp}J\vec{\kappa} }{{d}\tau}  =\frac{{d}\Delta_{||} }{{d}\tau}\vec{\kappa} + \Delta_{||}\frac{{d}\vec{\kappa} }{{d}\tau} + \frac{{d}\Delta_{\perp} }{{d}\tau}J\vec{\kappa} + \Delta_{\perp}\frac{{d}J\vec{\kappa} }{{d}\tau}.
\end{align}
Using the equation
\begin{equation}
    \frac{{d}\vec{\kappa} }{{d}\tau} = v\left(\frac{\vec{\nabla}q}{q},J\vec{\kappa}\right) J\vec{\kappa},
\end{equation}
we obtain 
\begin{align}
    \label{ddrdt1}& \frac{{d}\delta \vec{r}}{{d}\tau} =\left[\frac{{d}\Delta_{||} }{{d}\tau}- \Delta_{\perp}v\left(\frac{\vec{\nabla}q}{q},J\vec{\kappa}\right) \right]\vec{\kappa} + \left[\Delta_{||}v\left(\frac{\vec{\nabla}q}{q},J\vec{\kappa}\right)   + \frac{{d}\Delta_{\perp} }{{d}\tau}\right] J\vec{\kappa} .
\end{align}
Let us define two important small dimensionless parameters, which will be used in further calculations:
\begin{align}
& \Delta_0(\tau) := \frac{k_0(\tau,\mu',\nu')}{k_0(\tau,\mu,\nu)} - 1,
     &&\Delta_0|_{\tau = 0} \approx \frac{1}{k_0}\left[\frac{\partial k_0}{\partial\mu}\delta\mu + \frac{\partial k_0}{\partial\nu}\delta\nu\right],\\
& \Delta_{\alpha} (\tau) := \alpha  (\tau,\mu',\nu') - \alpha (\tau,\mu',\nu'), && \Delta_{\alpha}|_{\tau = 0} \approx \frac{\partial \alpha_0}{\partial\mu}\delta\mu + \frac{\partial \alpha_0}{\partial\nu}\delta\nu.
\end{align}
Now let's write out the derivative $\delta \vec{r}$ from other considerations:
\begin{align}
   \label{ddrdt} & \frac{{d}\delta \vec{r}}{{d}\tau} = \frac{{d} \vec{r}'}{{d}\tau} - \frac{{d} \vec{r}}{{d}\tau} = {v}' \vec{\kappa} ' - v\vec{\kappa}.
\end{align}
For ${v}'$ and $\vec{\kappa}'$ for sufficiently small $\delta\vec{r}$, $\Delta_{\alpha}$, $\Delta_0$the following is satisfied
\begin{align}
    & {v}' \approx {v} +  (\vec{\nabla} {v},\vec{\kappa} )\Delta_{||} +(\vec{\nabla} {v},J\vec{\kappa} )\Delta_{\perp} +  \frac{\partial v}{\partial k_0} k_0\Delta_0,\\
    & \vec{\kappa}' = \exp \left[\Delta_\alpha  J\right] \vec{\kappa} \approx \vec{\kappa} + \Delta_{\alpha} J\vec{\kappa}.
\end{align}
Then the equation \ref{ddrdt} has the form
\begin{align}
    \label{ddrdt2}& \frac{{d}\delta \vec{r}}{{d}\tau}=\left[ (\vec{\nabla}{v},\vec{\kappa} )\Delta_{||} + (\vec{\nabla} {v},J\vec{\kappa} )\Delta_{\perp} + \frac{\partial v}{\partial \kappa_0} k_0\Delta_0\right] \vec{\kappa} +  {v} \Delta_{\alpha}J\vec{\kappa}.
\end{align}
Comparing \ref{ddrdt1} and \ref{ddrdt2} and considering the orthogonality of $\vec{\kappa}$ and $J\vec{\kappa}$), we obtain a system of equations on the functions $\Delta_{|||}$ and $\Delta_{\perp}$:
\begin{align}
    & \frac{{d}\Delta_{||} }{{d}\tau} = v\left\{   \left(\frac{\vec{\nabla} {v}}{v},\vec{\kappa} \right)\Delta_{||} + \left(\frac{\vec{\nabla} {v}}{v}+\frac{\vec{\nabla}q}{q},J\vec{\kappa} \right)  \Delta_{\perp} +\frac{1}{v}\frac{\partial v}{\partial {\kappa_0}} k_0 \Delta_0 \right\}, \\
    &  \frac{{d}\Delta_{\perp} }{{d}\tau} =  v\left\{-\left(\frac{\vec{\nabla}q}{q},J\vec{\kappa}\right) \Delta_{||} + \Delta_{\alpha}\right\}.
\end{align}
Now let's write the equation in $\Delta\alpha$. From the definition we obtain
\begin{align}
    & \frac{{d}\Delta_{\alpha} }{{d}\tau} = \frac{{d} \alpha' }{{d}\tau} - \frac{{d} \alpha }{{d}\tau} = {v}'\left(\frac{\vec{\nabla}q'}{q'},J\vec{\kappa}'\right) - v\left(\frac{\vec{\nabla}q}{q},J\vec{\kappa}\right).
\end{align}
For ${v} '$ the expression was obtained earlier, now consider $\frac{\vec{\nabla}q'}{q'}$:
\begin{align}
    & \vec{\nabla}q' \approx \vec{\nabla}q + H(q)\vec{k} \Delta_{||} + H(q)J\vec{k} \Delta_{\perp} + \vec{\nabla}\frac{\partial q}{\partial k_0}k_0\Delta_0,\\
    & q' \approx q + (\vec{\nabla} {q},\vec{\kappa} )\Delta_{||} +(\vec{\nabla} {q},J\vec{\kappa} )\Delta_{\perp} +  \frac{\partial q}{\partial k_0} k_0 \Delta_0,\\
    \notag& \frac{\vec{\nabla}q'}{q'} \approx\frac{\vec{\nabla}q}{q}+ \frac{1}{q}\left[H(q)\vec{\kappa} - \left(\frac{\vec{\nabla} {q}}{q},\vec{\kappa} \right)\vec{\nabla}q\right]\Delta_{||}+\\ & +\frac{1}{q}\left[H(q)J\vec{\kappa} - \left(\frac{\vec{\nabla} {q}}{q},J\vec{\kappa} \right)\vec{\nabla}q\right]\Delta_{\perp}+ \frac{1}{q}\left[\vec{\nabla}\frac{\partial q}{\partial k_0} - \frac{1}{q}\frac{\partial q}{\partial k_0}\vec{\nabla} {q}\right]k_0\Delta_0.
\end{align}
Here $H(q)$ is the Hessian of the function $q$:
\begin{align}
    & H(q):= \left(\begin{array}{cc}
        \frac{\partial ^2 q}{\partial x ^2} & \frac{\partial ^2 q}{\partial x \partial y} \\
         \frac{\partial ^2 q}{\partial x \partial y}& \frac{\partial ^2 q}{\partial y ^2}
    \end{array}\right).
\end{align}
Then for the derivative $\Delta\alpha$ we obtain equation:
\begin{align*}
     \frac{{d}\Delta_{\alpha} }{{d}\tau} &\approx \left[{v} +  (\vec{\nabla} {v},\vec{\kappa} )\Delta_{||} + (\vec{\nabla} {v},J\vec{\kappa} )\Delta_{\perp} +   \frac{\partial v}{\partial k_0}k_0\Delta_0\right]\times \\ & \times \bigg(\frac{\vec{\nabla}q}{q}+ \frac{1}{q}\left[H(q)\vec{\kappa} - \left(\frac{\vec{\nabla} {q}}{q},\vec{\kappa} \right)\vec{\nabla}q\right]\Delta_{||} +\\&+\frac{1}{q}\left[H(q)J\vec{\kappa} - \left(\frac{\vec{\nabla} {q}}{q},J\vec{\kappa} \right)\vec{\nabla}q\right]\Delta_{\perp}+ \\ & +\frac{1}{q}\left[\vec{\nabla}\frac{\partial q}{\partial k_0} - \frac{1}{q}\frac{\partial q}{\partial k_0}\vec{\nabla} {q}\right]k_0\Delta_0 ,J\vec{\kappa} -\Delta_{\alpha} \vec{\kappa}\bigg)  -v\left(\frac{\vec{\nabla}q}{q},J\vec{\kappa}\right),
\end{align*}
which can be written as
\begin{align}
 \frac{{d}\Delta_{\alpha} }{{d}\tau} = v\left\{\mathfrak{F}_{||} \Delta_{||} +  \mathfrak{F}_{\perp} \Delta_{\perp}- \left(\frac{\vec{\nabla}q}{q},\vec{\kappa}\right)\Delta_{\alpha} + { \mathfrak{F}_{\kappa_0}} k_0 \Delta_0 \right\},  
\end{align}
where
\begin{align}
    & \mathfrak{F}_{||} := \left(\frac{\vec{\nabla}q}{q},J\vec{\kappa}\right) \left(\frac{\vec{\nabla}v}{v}-\frac{\vec{\nabla}q}{q},\vec{\kappa}\right) + \frac{1}{q}\left(H(q) \vec{\kappa},J\vec{\kappa}\right),\\
    & \mathfrak{F}_{\perp} := \left(\frac{\vec{\nabla}q}{q},J\vec{\kappa}\right) \left(\frac{\vec{\nabla}v}{v}-\frac{\vec{\nabla}q}{q},J\vec{\kappa}\right) + \frac{1}{q}\left(H(q) J\vec{\kappa},J\vec{\kappa}\right),\\
    & \mathfrak{F}_{k_0} := \left(\frac{\vec{\nabla}q}{q},J\vec{\kappa}\right) \left[\frac{1}{v}\frac{\partial v}{\partial k_0}-\frac{1}{q}\frac{\partial q}{\partial k_0}\right] + \frac{1}{q}\left(\vec{\nabla}\frac{\partial q}{\partial k_0} ,J\vec{\kappa}\right).
\end{align}
One can easily see that in all equations (on $\Delta_{||},\Delta_{\perp}, \Delta_{\alpha}$) the key role is played by the lo\-ga\-rif\-mic derivatives of functions with respect to the parameters $\vec{\kappa},J\vec{\kappa}, k_0$. By introducing the notations
\begin{align}
    & f_{||} := \left(\frac{\vec{\nabla f}}{f},\vec{\kappa}\right),\quad  f_{\perp} := \left(\frac{\vec{\nabla f}}{f},J\vec{\kappa}\right),\quad f_{0} := \frac{1}{f}\frac{\partial f}{\partial k_0},
\end{align}
we obtain a simplified form of the equations
\begin{align}
     \frac{d \Delta_{||}}{d\tau} =& \,v\left\{ v_{||} \Delta_{||} + (v_{\perp}+q_{\perp})\Delta_{\perp} + v_0k_0\Delta_0 \right\},\\
     \frac{d \Delta_{\perp}}{d\tau} =&\, v\left\{ - q_{\perp}\Delta_{||} +\Delta_{\alpha} \right\},\\
   \notag  \frac{d \Delta_{\alpha}}{d\tau} =& \,v\bigg\{\left[q_{\perp} (v_{||}-q_{||}) +\left(\frac{1}{q}H(q) \vec{\kappa},J\vec{\kappa}\right)\right]\Delta_{||} + \\ \notag& +\left[q_{\perp} (v_{\perp}-q_{\perp}) +\left(\frac{1}{q}H(q) J\vec{\kappa},J\vec{\kappa}\right)\right]\Delta_{\perp}  -\\& - q_{||}\Delta_{\alpha} + \left[q_{\perp} (v_{0}-q_{0}) +\left(\frac{1}{q}\vec{\nabla}\frac{\partial q}{\partial k_0} ,J\vec{\kappa}\right)\right]k_0\Delta_{0}\bigg\},\\
    \frac{d \Delta_{0}}{d\tau} = & \, 0.
\end{align}
The initial data for $\Delta_\alpha$ and $\Delta_0$ have been described above, and for $\Delta_{|||}$ and $\Delta_{\perp}$ are defined similarly:
\begin{align}
    & \Delta_{||}|_{\tau = 0 } = \left(\frac{\partial \vec{r}_0}{\partial \mu} \delta\mu + \frac{\partial \vec{r}_0}{\partial \nu} \delta\nu, \vec{\kappa} (\alpha_0 (\mu,\nu))\right),\\
        & \Delta_{\perp}|_{\tau = 0 } = \left(\frac{\partial \vec{r}_0}{\partial \mu} \delta\mu + \frac{\partial \vec{r}_0}{\partial \nu} \delta\nu, J\vec{\kappa} (\alpha_0 (\mu,\nu))\right).
\end{align}
The equations can be written as a system
\begin{align}
   \label{system 1} & \frac{d}{d\tau}\mathbf{\Delta} = v A\mathbf{\Delta},\quad \mathbf{\Delta}:= \left(\begin{array}{c}
         \Delta_{||}  \\
         \Delta_{\perp}\\
         \Delta_{\alpha}\\
         \Delta_0
    \end{array}\right),\quad \mathbf{\Delta}|_{\tau = 0} = \mathbf{\Delta}_{\mu} \delta\mu + \mathbf{\Delta}_{\nu} \delta\nu  = \\& \quad= \left(\begin{array}{c}
        \left(\frac{\partial \vec{r}_0}{\partial \mu} , \vec{\kappa} (\alpha_0)\right)  \\
         \left(\frac{\partial \vec{r}_0}{\partial \mu}, J\vec{\kappa} (\alpha_0)\right)\\\frac{\partial \alpha_0}{\partial\mu}\\
         \frac{1}{k_0}\frac{\partial k_0}{\partial\mu}    \end{array}\right) \delta\mu + \left(\begin{array}{c}
        \left(\frac{\partial \vec{r}_0}{\partial \nu} , \vec{\kappa} (\alpha_0)\right)  \\
         \left(\frac{\partial \vec{r}_0}{\partial \nu}, J\vec{\kappa} (\alpha_0)\right)\\\frac{\partial \alpha_0}{\partial\nu}\\
         \frac{1}{k_0}\frac{\partial k_0}{\partial\nu}    \end{array}\right)\delta\nu,
\end{align}
where $A$ is a $4\times4$ matrix-function the elements of which are defined by the equations above. It is also convenient to consider the fundamental matrix solution $\mathcal{M}$:
\begin{align}
    \label{system 2}&  \frac{{d} }{{d}\tau} \mathcal{M} = v{A}\mathcal{M},\quad \mathcal{M}|_{\tau=0} = I.
\end{align}
In terms of the matrix $\mathcal{M}(\tau,\mu,\nu)$, the derivatives of the vector $\vec{r}$ and the angle $\alpha$ with respect to the parameters $\mu$ and $\nu$ are approximated:
\begin{align*}
    & \frac{\partial \vec{r}}{\partial \mu}  \approx (\mathcal{M}\mathbf{\Delta}_\mu)_{1} \vec{\kappa} + (\mathcal{M}\mathbf{\Delta}_\mu)_{2} J\vec{\kappa}
    && \frac{\partial \vec{r}}{\partial \nu}  \approx (\mathcal{M}\mathbf{\Delta}_\nu)_{1} \vec{\kappa} + (\mathcal{M}\mathbf{\Delta}_\nu)_{2} J\vec{\kappa},\\
    & \frac{\partial \alpha}{\partial \mu} \approx (\mathcal{M}\mathbf{\Delta}_\mu)_{3},
    && \frac{\partial \alpha}{\partial \nu} \approx (\mathcal{M}\mathbf{\Delta}_\nu)_{3}.
\end{align*}
Thus, we can consider that the solutions of the systems \ref{system 1} or \ref{system 2} determine the derivatives of $\vec{r}$ and $\alpha$ at the ray coordinates $\mu$ and $\nu$ along the ray.
\subsection{Jacobi matrix}
Now we will show that the solution of the systems \ref{system 1},\ref{system 2} also allows us to determine the location and parameters of space-time caustics. This requires the ability to compute the Jacobian of the transition from coordinates $\rho,\vec{r}$ to ray coordinates $\tau,\mu,\nu$. Let us recall that the Jacobian is defined by the equality
\begin{align}
    & D(\tau,\mu,\nu) = \det \mathcal{J},\quad \mathcal{J}= \left(\begin{array}{ccc}
        \frac{\partial \rho}{\partial \tau} & \frac{\partial \rho}{\partial \mu} &\frac{\partial \rho}{\partial \nu}\\
        \frac{\partial \vec{r}}{\partial \tau} & \frac{\partial \vec{r}}{\partial \mu} &\frac{\partial \vec{r}}{\partial \nu}\\
    \end{array}\right) = \left(\begin{array}{ccc}
        1 & \frac{\partial \rho_0}{\partial \mu} &\frac{\partial \rho_0}{\partial \nu}\\
        \frac{\partial \vec{r}}{\partial \tau} & \frac{\partial \vec{r}}{\partial \mu} &\frac{\partial \vec{r}}{\partial \nu}\\
    \end{array}\right)
\end{align}
Let us express the Jacobi matrix in terms of the matrix $\mathcal{M}$:
\begin{align}
    & \mathcal{J} = \left(\begin{array}{ccc}
1 & \frac{\partial \rho_0}{\partial \mu} &\frac{\partial \rho_0}{\partial \nu}\\
        v \vec{\kappa} & (\mathcal{M}\mathbf{\Delta}_\mu)_{1} \vec{\kappa} + (\mathcal{M}\mathbf{\Delta}_\mu)_{2} J\vec{\kappa} &(\mathcal{M}\mathbf{\Delta}_\nu)_{1} \vec{\kappa} + (\mathcal{M}\mathbf{\Delta}_\nu)_{2} J\vec{\kappa}\\
    \end{array}\right).
\end{align}
Then for the Jacobian, using the equality 
\begin{align}
    & \det(\vec{a}\,\,\vec{b}) = (J\vec{a},\vec{b}),
\end{align}
we obtain
\begin{align}
     D(\tau,\mu,\nu) &= \left[(\mathcal{M}\mathbf{\Delta}_\mu)_{1}(\mathcal{M}\mathbf{\Delta}_\nu)_{1} - (\mathcal{M}\mathbf{\Delta}_\mu)_{2}(\mathcal{M}\mathbf{\Delta}_\nu)_{2}\right] + \\&+v\left[\frac{\partial \rho_0}{\partial \nu}(\mathcal{M}\mathbf{\Delta}_\mu)_{2} - \frac{\partial \rho_0}{\partial \mu}(\mathcal{M}\mathbf{\Delta}_\nu)_{2} \right]
\end{align}
We also note that 
\begin{align}
    & D_0(\mu,\nu):=D|_{\tau = 0} =\left|\begin{array}{ccc}
        1 & \frac{\partial \rho_0}{\partial \mu} &\frac{\partial \rho_0}{\partial \nu}\\
        v\vec{\kappa}(\alpha_0) & \frac{\partial \vec{r}_0}{\partial \mu} &\frac{\partial \vec{r}_0}{\partial \nu}
    \end{array}\right|.
\end{align}
The condition $D_0\neq 0$ ensures unambiguity of ray coordinates in some neighborhood of the initial surface. 
We will call space-time caustics those points in space $\rho(\tau,\mu,\nu),\vec{r}(\tau,\mu,\nu)$ for which the Jacobian $D(\tau,\mu,\nu)$ turns into 0. 
\subsection{Observation in the neighborhood of a point}
Let us introduce the notations
\begin{align}
& \mathcal{R} = \left(\rho,x,y\right)^T,\quad \mathcal{T}:=\left(\tau,\mu,\nu\right)^T
\end{align}
and consider the following problem formulation. Consider some point $\mathcal{R}$ in space-time, which we will call the observation point. We will assume that it is known that there exists a space-time ray passing through this point, and therefore $\mathcal{R} = \mathcal{R} (\mathcal{T})$. The ray arrives at the point with the parameters
\begin{align}
 & \rho = \tau +  \rho_{0}(\mu,\nu), &&\vec{r} = \vec{r} (\tau,\mu,\nu)\\
  & k_0 = k_0(\mu,\nu), && \alpha = \alpha(\tau,\mu,\nu).
\end{align}
The variation of the ray with respect to the ray parameters in the neighborhood of the observation point is described in the form:
\begin{align}
     & \mathcal{R}(\mathcal{T} + \delta\mathcal{T}) \approx \mathcal{R} + \mathcal{J}(\mathcal{T})\delta\mathcal{T}, 
\end{align}
where $\mathcal{J}$ is the $3\times3$ Jacobi matrix of the transition to ray coordinates. We will be interested in the representations for various field parameters in the neighborhood of this spacetime point. Considering all functions to be smooth and the observation point to be far from space-time caustics, we can write out relations for the parameter variations and the gradients in different coordinates:
\begin{align}
    & \delta\mathcal{R} = \mathcal{J}(\mathcal{T})\delta\mathcal{T},&&\delta\mathcal{T} = \mathcal{J}^{-1}(\mathcal{T})\delta\mathcal{R},\\
    & \hat{\nabla}_{\mathcal{R}} f= (\mathcal{J}^*(\mathcal{T}))^{-1}\hat{\nabla}_{\mathcal{T}} f, && \hat{\nabla}_{\mathcal{T}} f= \mathcal{J}^*(\mathcal{T})\hat{\nabla}_{\mathcal{R}} f,
\end{align}
where $f$ is an arbitrary function, and $\hat{\nabla}_{\mathcal{R}}$ and $\hat{\nabla}_{\mathcal{T}}$ are three-dimensional (space-time) gradients in regular and ray coordinates, respectively.
\subsection{Conditions on the initial data}
The important thing for the use of the apparatus described above is the coherence of the initial data - the functions at $\tau = 0$ cannot be defined arbitrarily. Recall that there are functions determined on the surface given by ray coordinates $\mu,\nu$
\begin{align}
    & \rho_0(\mu,\nu), \vec{r}_0(\mu,\nu), k_0(\mu,\nu), \alpha_0(\mu,\nu), \phi_0(\mu,\nu), A(\mu,\nu).
\end{align}
The nondegeneracy of the Jacobi matrix is required for correctness of the ray approximation at the initial point
\begin{align}
    & \mathcal{J}_0 = \left(\begin{array}{ccc}
        1 & \frac{\partial \rho_0}{\partial \mu} &\frac{\partial \rho_0}{\partial \nu}\\
        v\vec{\kappa}(\alpha_0) & \frac{\partial \vec{r}_0}{\partial \mu} &\frac{\partial \vec{r}_0}{\partial \nu}\end{array}\right),\quad\det \mathcal{J}_0 \neq 0.
\end{align}
The following equations are satisfied according to the definition
\begin{align}
    & k_0(\mu,\nu) =k_0(\rho_0(\mu,\nu),\vec{r}_0(\mu,\nu)) = \frac{\partial \phi(\tau,\mu,\nu)}{\partial\rho(\tau,\mu,\nu)}\bigg|_{\tau = 0},\\
    &\vec{\kappa}(\alpha_0(\mu,\nu)) = \vec{\kappa}(\alpha_0(\rho_0(\mu,\nu),\vec{r}_0(\mu,\nu))) = \frac{\partial \phi(\tau,\mu,\nu)}{\partial\vec{r}(\tau,\mu,\nu)}\bigg|_{\tau = 0}.
\end{align}
Let us apply the considerations described earlier for the case of the observation point 
\begin{align}    &\mathcal{R}_{obs} = (\rho_0(\mu,\nu),\vec{r}_0(\mu,\nu))^T.
\end{align}
 and the corresponding point
\begin{align}
    &\mathcal{T}_{obs} = (0,\mu,\nu)^T.
\end{align}
Then the following equality holds
\begin{align}
    & \hat{\nabla}_{\mathcal{T}} \phi(\mathcal{T}_{obs}) = (\mathcal{J}_0^*)\hat{\nabla}_{\mathcal{R}} \phi(\mathcal{R}_{obs}).
\end{align}
Hence we get the requirements for the function $\phi_0$:
\begin{align}
    & \left(\begin{array}{c}
         \frac{\partial\phi}{\partial\tau} \\
         \frac{\partial\phi}{\partial\mu}\\
         \frac{\partial\phi}{\partial\nu}
    \end{array}\right)\bigg|_{\tau = 0} = \mathcal{J}_0^*\left(\begin{array}{c}
         k_0(\mu,\nu) \\
         q(\vec{r}_0(\mu,\nu),k_0(\mu,\nu))\cos \alpha_0(\mu,\nu)\\
         q(\vec{r}_0(\mu,\nu),k_0(\mu,\nu))\sin \alpha_0(\mu,\nu)
    \end{array}\right)
\end{align}
Knowing that
\begin{align}
& \frac{\partial\phi}{\partial\tau}|_{\tau=0} = k_0(\mu,\nu) + q(\vec{r}_0(\mu,\nu),k_0(\mu,\nu)) v(\vec{r}_0(\mu,\nu),k_0(\mu,\nu)),
\end{align}
we obtain that the equation above is a first order partial derivative equation on the function $\phi_0(\mu,\nu)$. The amplitude, on the other hand, does not have such strict restrictions on the coherence of the initial data, so the only limitation for its setting is the requirement of the smooth and slow (relative to the phase) variation with respect to the initial parameters. 
\subsection{Fronts}
Let $f(\tau,\mu,\nu)$ be some function defined along the rays. We define the surface of constant value of this function by the equality:
\begin{align}
    & \mathfrak{F}[f,c]:=\left\{ (\rho,x,y)^T = \mathcal{R}(\mathcal{T})|\,\, f(\mathcal{T}) = c\right\}\subset\mathbb{R}^3.
\end{align}
We will be interested not only in such a surface itself, but also in its projection onto the plane $(x,y)$. For this purpose, let us define the orthogonal projector $P_{xy}$ by the following equality
\begin{align}
    & R_{xy}:\mathbb{R}^3\to\mathbb{R}^2,\quad P_{xy}\mathcal{R} : = (x,y)^T,\,\,\mathcal{R}=(\rho,x,y)^T
\end{align}
Let us fix the point $\mathcal{R}_0 = (\rho_0,x_0,y_0)^T$, which corresponds to a point in ray coordinates $\mathcal{T}_0 = (\tau_0, \mu_0,\nu_0)^T$, and consider its $\epsilon-$neighborhood $U_{\epsilon}(\mathcal{R}_0)\subset\mathbb{R}^3$. Then the \textit{local surface} and \textit{local front} of the function $f$ in the neighborhood of the point $\mathcal{R}_0$ we will call the sets
\begin{align}
    &\mathfrak{F}_{\epsilon}[f,\mathcal{R}_0] := \mathfrak{F}[f,f(\mathcal{T}_0)]\cap U_{\epsilon}(\mathcal{R}_0)\\
    & \mathfrak{f}_{\epsilon}[f,\mathcal{R}_0]:=P_{xy}\mathfrak{F}_{\epsilon}[f,\mathcal{R}_0].
\end{align}
We will be interested in the case when the local surface turns out to be two-dimensional and the front turns out to be a line. It is easy to see that the (three-dimensional) normal to the surface is defined by the equation
\begin{align}
    &\hat{n}_f=\hat{\nabla}_{\mathcal{R}}f = (\mathcal{J}^*)^{-1}\hat{\nabla}_{\mathcal{T}}f\neq 0,
\end{align}
and the front normal, in its turn 
\begin{align}
    & \vec{n}_f= P_{xy}\hat{n}_f =P_{xy}\hat{\nabla}_{\mathcal{R}}f =P_{xy} (\mathcal{J}^*)^{-1}\hat{\nabla}_{\mathcal{T}}f\neq 0.
\end{align}
If thus defined normals turn out to be nonzero and correctly defined for any neighborhood of the initial point $\mathcal{R}_0$, the corresponding sets will be called \textit{global surface} and \textit{global front} of the function $f$ (or, shortening, $f$-surface and $f$-front):
\begin{align}
    & \mathfrak{F}[f,\mathcal{R}_0] := \mathfrak{F}_{\infty}[f,\mathcal{R}_0], &&\mathfrak{f}[f,\mathcal{R}_0]:=\mathfrak{f}_{\infty}[f,\mathcal{R}_0] .
\end{align}

\section{Observable parameters}
Let us use the definitions of $f$- fronts and $f$-surfaces for a more detailed study of the signal propagation. For this purpose, let us consider most typical examples of $f$-functions.
\subsection{Phase front}
Let the phase $\phi$ be chosen as $f$:
\begin{align}
    & \phi(\tau,\mu,\nu) = \phi_0(\mu,\nu) + k_0(\mu,\nu) \tau + \int_{0}^{\tau} q(\tau',\mu,\nu)v(\tau',\mu,\nu)d\tau'.
\end{align}
From the definition, we can see that the gradient in ray coordinates is explicitly computable:
\begin{align}
 &\hat{\nabla}_{\mathcal{T}}\phi = \left(k_0 + qv,\frac{\partial\phi}{\partial\mu},\frac{\partial\phi}{\partial\nu}\right)^T,\\
\notag&\frac{\partial \phi}{\partial \xi} = \frac{\partial \phi_0}{\partial  \xi} + \frac{\partial k_0}{\partial  \xi}\left[\tau + \int_0^{\tau}qv\left\{\frac{1}{q}\frac{\partial q}{\partial  k_0} + \frac{1}{v}\frac{\partial v}{\partial  k_0}\right\} d\tau'\right] +\\ 
 &+ \int_0^{\tau}  qv\left\{\left(\frac{\vec{\nabla }q}{q}, \frac{\partial\vec{r}}{\partial \xi}\right) + \left(\frac{\vec{\nabla }v}{v}, \frac{\partial\vec{r}}{\partial \xi}\right)\right\}d\tau', \quad \xi = \mu,\nu.
\end{align}
Then for the normal $\hat{n}_{\phi}$ there exists the following interpretation 
\begin{align}
    &\hat{n}_{\phi} = (\mathcal{J}^*)^{-1}\hat{\nabla}_{\mathcal{T}}\phi = \left(\mathbf{k}_0,\vec{\mathbf{k}}\right)^T ,
\end{align}
where $\vec{\mathbf{k}}$ is the \textit{observed} wave vector and $\mathbf{k}_0$ is the \textit{observed} frequency. Their difference from $q\vec{\kappa}$ and $k_0$ lies in the fact that they reflect the propagation of the signal as a single object, rather than being parameters of a particular space-time ray. In this case, at a fixed spatial point $(x_0,y_0)$ function \begin{align}
    & \mathbf{k}_0 (\rho) := \mathbf{k}_0(\rho, x_0,y_0)
\end{align}
 describes the frequency modulation of the observed signal. 
 \subsection{$\tau$-front and $s-$front}
Suppose that the coordinate $\tau$ which characterizes the time of motion along the ray to the observation point is chosen as $f$. Then \begin{align}
 &\hat{\nabla}_{\mathcal{T}}\tau = \left(1,0,0\right)^T,
\end{align}
and the $\tau$- front represents the set of points at time $\tau$ after their radiation. Now let us consider as $f$ a function \begin{align}
    & s(\tau,\mu,\nu) := \int_{0}^{\tau} v(\tau',\mu,\nu) d\tau',
\end{align}
characterizing the length of the path from the initial point to the observation point. Then 
\begin{align}
 &\hat{\nabla}_{\mathcal{T}}s = \left(v,\frac{\partial s}{\partial \mu},\frac{\partial s}{\partial \nu}\right)^T,\\& \frac{\partial s}{\partial \xi} =  \int_{0}^{\tau} \left\{\left(\vec{\nabla}v,\frac{\partial \vec{r}}{\partial\xi}\right) + \frac{\partial v}{\partial k_0}\frac{\partial k_0}{\partial \xi} \right\}d\tau',\quad \xi=\mu,\nu.
\end{align}
One can easily see that in the case of a constant group velocity the $\tau$front and the $s-$front coincide, but in the case of an inhomogeneous medium these sets turn out to be, generally speaking, different. If the propagation of rays occurs far from caustics, an asymptotic expression for the amplitude can be derived:
\begin{align}
    & A(\tau,\mu,\nu)\sim A(0,\mu,\nu) \frac{1}{\sqrt{s(\tau)}},
\end{align}
which allows us to speak of the $s-$-front as a \textit{amplitude} front. However, to consider a real amplitude front ($A$-front), one needs to calculate the derivatives of the Jacobian of the transition into ray coordinates, which is beyond the scope of considering the system of equations in the first approximation. 

\subsection{Multiple rays observations}
Suppose that $N$ space-time rays with significantly different ray coordinates pass through the observation point $\mathcal{R}_0$:
\begin{align}
    & \mathcal{R}(\mathcal{T}_1) = \dots = \mathcal{R}(\mathcal{T}_N) = \mathcal{R}_0.
\end{align}
Then the total field at a point is given by the equality 
\begin{align}
    & U (\mathcal{R}_0) \approx \sum_{j=1}^N A_j(\mathcal{T}_j) \exp{\left\{i\varepsilon^{-1}\phi_j(\mathcal{T}_j)\right\}}.
\end{align}
Considering that the amplitudes in the neighborhood of the observation point change slower than the phases, we obtain the description of the total field variation
\begin{align}
    \label{u full 1}&U (\mathcal{R}_0 + \delta\mathcal{R})\approx\sum_{j=1}^N A_j \exp{\left\{i\varepsilon^{-1}\phi_j(\mathcal{T}_j) \left[ 1 +\frac{((\mathcal{J}_{j}^{*})^{-1}\hat{\nabla}_{\mathcal{T}}\phi_j,\delta\mathcal{R})}{\phi_j(\mathcal{T}_j)}\right]\right\}}.
\end{align}
Then
\begin{align}
    \label{u full 2}&\frac{\partial U}{\partial \mathcal{R}} \approx\sum_{j=1}^N i\varepsilon^{-1} A_j\exp{\left\{i\varepsilon^{-1}\phi_j(\mathcal{T}_j)\right\}}  {(\mathcal{J}_{j}^{*})^{-1}\hat{\nabla}_{\mathcal{T}}\phi_j} .
\end{align}
Thus, in the case multiple rays it is not possible to speak about the instantaneous frequency and wave vector, as well as about the separated amplitude function - the signal has a complex structure that depends significantly on the parameters of the incoming rays. However, in each specific case, rough estimations of the expressions \ref{u full 1} and \ref{u full 2} are possible by using the techniques described above. 
\section{Final commentaries}
Note that despite the scaling transformation $c_0 T,X,Y,Z \to \tau,x,y,z$ made at the very beginning, which depends on the choice of $\varepsilon$, the actual appearance of this transformation is only in the correct definition of the relation between the vertical coordinate and all others. Due to the fact that the influence of this coordinate only manifests itself in the definition and properties of the function $q$, in all further reasoning we can perform the inverse transformation and consider that \begin{align}
    & \tau = c_0 T,\quad x =X,\quad y = Y, \quad \varepsilon = 1
\end{align}
which corresponds to an alternative (physical) approach to the construction of the asymptotic decomposition \cite{kravtsov1990geometrical}. 
\bibliographystyle{plain}
\bibliography{main}

\begin{thebibliography}{10}

\bibitem{babich1998space}
Vasili~M. Babich, Ivan~A. Molotkov, and Vladimir~S. Buldyrev.
\newblock {\em The space-time ray method}.
\newblock Cambridge Univ. Press, 1998.

\bibitem{babich1980space}
Vasilii~Mikhailovich Babich and Natal'ya~Sergeevna Grigor'eva.
\newblock Space-time ray method of the calculation of waves in slightly inhomogeneous layered medium.
\newblock {\em Zapiski Nauchnykh Seminarov POMI}, 99:5--18, 1980.

\bibitem{babich1984complex}
V.M. Babich and V.V. Ulin.
\newblock Complex space-time ray method and “quasiphotons”.
\newblock {\em Journal of Soviet Mathematics}, 24:269--273, 1984.

\bibitem{bergman2005generalized}
D.R. Bergman.
\newblock Generalized space-time paraxial acoustic ray tracing.
\newblock {\em Waves in Random and Complex Media}, 15(4):417--435, 2005.

\bibitem{bor1985group}
Zs. Bor and B.~Racz.
\newblock Group velocity dispersion in prisms and its application to pulse compression and travelling-wave excitation.
\newblock {\em Optics communications}, 54(3):165--170, 1985.

\bibitem{burridge2005horizontal}
Robert Burridge and Henry Weinberg.
\newblock Horizontal rays and vertical modes.
\newblock {\em Wave propagation and underwater acoustics}, pages 86--152, 2005.

\bibitem{vcerveny1982space}
Vlastislav {\v{C}}erven{\`y}, I.A. Molotkov, and Ivan P{\v{s}}en{\v{c}}{\'\i}k.
\newblock Space-time ray method for seismic wave fields.
\newblock {\em Studia Geophysica et Geodaetica}, 26(4):342--351, 1982.

\bibitem{connor1974complex}
Kenneth~Allen Connor and L.B. Felsen.
\newblock Complex space-time rays and their application to pulse propagation in lossy dispersive media.
\newblock {\em Proceedings of the IEEE}, 62(11):1586--1598, 1974.

\bibitem{etter2018underwater}
Paul~C. Etter.
\newblock {\em Underwater acoustic modeling and simulation}.
\newblock CRC press, 2018.

\bibitem{felsen1981hybrid}
L.B. Felsen.
\newblock Hybrid ray-mode fields in inhomogeneous waveguides and ducts.
\newblock {\em The Journal of the Acoustical Society of America}, 69(2):352--361, 1981.

\bibitem{hebling1996derivation}
J.~Hebling.
\newblock Derivation of the pulse front tilt caused by angular dispersion.
\newblock {\em Optical and Quantum Electronics}, 28:1759--1763, 1996.

\bibitem{hebling2008generation}
J{\'a}nos Hebling, Ka-Lo Yeh, Matthias~C Hoffmann, Bal{\'a}zs Bartal, and Keith~A Nelson.
\newblock Generation of high-power terahertz pulses by tilted-pulse-front excitation and their application possibilities.
\newblock {\em JOSA B}, 25(7):B6--B19, 2008.

\bibitem{katsnelson2012fundamentals}
Boris Katsnelson, Valery Petnikov, and James Lynch.
\newblock {\em Fundamentals of shallow water acoustics}.
\newblock Springer Science \& Business Media, 2012.

\bibitem{katsnelson2018variability}
Boris~G. Katsnelson, Valery~A. Grigorev, and James~F. Lynch.
\newblock Variability of phase and amplitude fronts due to horizontal refraction in shallow water.
\newblock {\em The Journal of the Acoustical Society of America}, 143(1):193--201, 2018.

\bibitem{katsnel2012space}
B.G. Katsnel’Son and A.~Yu. Malykhin.
\newblock Space-time sound field interference in the horizontal plane in a coastal slope region.
\newblock {\em Acoustical Physics}, 58:301--307, 2012.

\bibitem{kirpichnikova1983reflection}
Natal'ya~Yakovlevna Kirpichnikova and Mikhail~Mikhailovich Popov.
\newblock Reflection of space-time ray amplitudes from arbitrary moving boundary.
\newblock {\em Zapiski Nauchnykh Seminarov POMI}, 128:72--88, 1983.

\bibitem{kravtsov2012caustics}
Yu.~A. Kravtsov and Yu.~I. Orlov.
\newblock {\em Caustics, catastrophes and wave fields}, volume~15.
\newblock Springer Science \& Business Media, 2012.

\bibitem{kravtsov1990geometrical}
Yury~A. Kravtsov and Yuri~Ilich Orlov.
\newblock {\em Geometrical optics of inhomogeneous media}, volume~38.
\newblock Springer, 1990.

\bibitem{osvay2004angular}
K{\'a}roly Osvay, Attila~P. Kov{\'a}cs, Zsuzsanna Heiner, G{\'a}bor Kurdi, J{\'o}zsef Klebniczki, and M{\'a}rta Csat{\'a}ri.
\newblock Angular dispersion and temporal change of femtosecond pulses from misaligned pulse compressors.
\newblock {\em IEEE Journal of selected topics in quantum electronics}, 10(1):213--220, 2004.

\bibitem{topp1975group}
Michael~R. Topp and Gail~C. Orner.
\newblock Group dispersion effects in picosecond spectroscopy.
\newblock {\em Optics Communications}, 13(3):276--281, 1975.

\bibitem{weinberg1974horizontal}
Henry Weinberg and Robert Burridge.
\newblock Horizontal ray theory for ocean acoustics.
\newblock {\em The Journal of the Acoustical Society of America}, 55(1):63--79, 1974.

\end{thebibliography}

\end{document}